\documentclass[aps,amsfonts,nofootinbib,article,showpacs,superscriptaddress]{revtex4}
\usepackage{graphicx}
\usepackage{amsmath}
\usepackage{amsfonts}
\usepackage{amssymb}
\usepackage{mathrsfs}

\def\openone{\leavevmode\hbox{\small1\kern-3.8pt\normalsize1}}
\def\N{\leavevmode\hbox{ Z \kern-8 pt\normalsize{Z}}}
\def\openone{\leavevmode\hbox{\small1\kern-3.8pt\normalsize1}}
\def\openJ{\leavevmode\hbox{J \kern-9.5pt\normalsize J}}
\def\openS{\leavevmode\hbox{ S \kern-9.3pt\normalsize S}}
\newcommand{\bb}{\begin{equation}}
\newcommand{\ee}{\end{equation}}
\newcommand{\eqb}{\begin{eqnarray}}
\newcommand{\eqf}{\end{eqnarray}}

\usepackage{color}

\begin{document}

\title{Non--geodesic circular motion of massive spinning test bodies}

\author{Sergio A. Hojman}
\email{sergio.hojman@uai.cl}
\affiliation{Departamento de Ciencias, Facultad de Artes Liberales,
Universidad Adolfo Ib\'a\~nez, Santiago, Chile.}
\affiliation{Centro de Investigaci\'on en Matem\'aticas, A.C., Unidad M\'erida; Yuc, M\'exico. }
\affiliation{Departamento de F\'{\i}sica, Facultad de Ciencias, Universidad de Chile,
Santiago, Chile.}
\affiliation{Centro de Recursos Educativos Avanzados,
CREA, Santiago, Chile.}
\author{Felipe A. Asenjo}
\email{felipe.asenjo@uai.cl}
\affiliation{Facultad de Ingenier\'{\i}a y Ciencias,
Universidad Adolfo Ib\'a\~nez, Santiago, Chile.}

\begin{abstract}
Recent interest on studying possible violations of the Equivalence Principle has led to the development of space satellite missions testing it for bodies moving on circular orbits around Earth. This experiment establishes that the validity of the Equivalence Principle is independent of the composition of bodies.  However, the internal degrees of freedom of the bodies (such as spin) were not taken into account.
In this work, it is shown exactly that the circular orbit motion of test bodies does present a departure from geodesic motion when spin effects are not negligible. Using a Lagrangian theory for spinning massive bodies, an exact solution for their circular motion  is found showing that the non--geodesic behavior manifests through different tangential velocities of the test bodies, depending on the orientation of its spin  with respect to the total angular momentum of the satellite. Besides, for circular orbits, spinning test bodies present no tangential acceleration. We estimate the difference of the two possible tangential velocities  for the case of circular motion of spinning test bodies orbiting  Earth.
\end{abstract}


\maketitle

\section{Introduction}

 The Equivalence Principle (EP) is one of the cornerstones of General Relativity. Among all the different possible ways in which it has been stated, one of its simplest form (called its weak form) establishes that all bodies fall with the same acceleration in a given gravitational field \cite{weinb},  implying the equivalence between gravitational and inertial masses.
 Another  precise form to enunciate it is that all bodies moving under the influence of gravitational forces only follow geodesics \cite{weinb}. The EP only applies in a region of spacetime small enough to neglect the inhomogeneities of  gravitational fields \cite{weinb}.

In order to determine experimentally  the validity of the EP, some experiments have been carried out recently in different settings \cite{microscope,UFF,verena,tarallo,pulsar1EP,pulsar2EP}. They consider the trajectories of massive composed falling bodies, measuring their accelerations, and determining whether (or not) the E\"otv\"os ratio parameter that characterizes the falling is non-zero. The E\"otv\"os parameter $\Delta$ measures the relative difference between accelerations  for falling test bodies, and according to the EP, it should vanish.

These experiments run from atomic to celestial scales. For instance, in Ref.~\cite{UFF}, $^{87}$Rb atoms were studied in a vertical free--falling configuration, where the cluster spin
was vertically aligned, pointing either up or down. This experiment determined that  $\Delta\sim 10^{-7}$, establishing that the experimental results were not in agreement with any of the considered theoretical models for spin-curvature and spin-torsion couplings devepoled in Refs.~\cite{spgr1,spgr2,spgr3}. Other experiments testing EP with atoms  have been perfomed in Refs.~\cite{verena,tarallo}. The kind of experiment  performed by the MICROSCOPE satellite (MS) mission \cite{microscope} is different. The aim of this mission was to measure the forces required to maintain two cylindrical test massive bodies in the same circular orbit around Earth. Bodies with the same and different compositions showed no difference on their trajectory behavior, finding an  E\"otv\"os ratio of the order $\Delta \sim 10^{-15}$. The MS mission was focused in determining if the atomic composition of massive bodies can produce any violation of EP, and their findings have a strong indication that it does not. Even tough, recent observations have helped to probe the validity of EP at galactic scales \cite{pulsar1EP,pulsar2EP}.

However, these experiments do not consider any internal degree of freedom of the test bodies, such as spin.  In general, it is well--known that spin introduces tidal forces that deviate any massive free--falling spinning body from a geodesic. Therefore, any spinning massive particle does not follow geodesics \cite{barker}.
The pioneering works of Mathisson \cite{math}  and
Papapetrou \cite{papa}  showed that the equations of motion for spinning massive particles are non--geodesic, deriving them
as limiting cases of rotating fluids moving in gravitational fields. The  Mathisson-Papapetrou equations (MPE)
have been used to obtain several exact solutions (see for example Refs.~\cite{mashhoon,plya1,waldo,plya2,oseme}). However, the MPE present several problems in the description of spinning massive particles. For example, they are dynamical equations of third order that do not preserve the square root of the Casimir operator of the Poincar\'e group $P_\mu P^\mu$ (formed by the momentum $P^\mu$),  among others. These difficulties are analyzed below \cite{hojmanComment}.

Action approaches can be used to describe the dynamics of spinning massive particles from first principles \cite{Steinhoff}.
 In this work, we use the Lagrangian theory developed in Res.~\cite{hojman1,hr,hojman2a,hojman2,hojman3,ha,hojmanUFF,hk,zha,abk,ahkz,nz,zgwyl}, which also allow us to avoid the several  issues of MPE. This theory can be derived rigously from first principles allowing the proper treatment of the crucial lack of parallelism between velocity
and momentum, which otherwise cannot be obtained as the canonical momentum cannot be appropriately defined (as in MPE). This Lagrangian theory has been used to study spinning massive particles (tops) in different contexts and gravitational fields \cite{hojman2a,hojman2,ha,hk,zha,abk,ahkz,nz,zgwyl}, always finding new effects on the dynamics associated to the non--geodesics motion of tops due to spin--gravity coupling.
Moreover,  in Ref.~\cite{hojmanUFF} was shown that this Lagrangian model for tops matches the experimental conclusions of  Ref.~\cite{UFF}. The Lagrangian  theory exactly predicts the results of $^{87}$Rb atoms experiment \cite{UFF}, rigourously showing that in tops in a vertical free--falling trajectory with spins  aligned (with the trajectory), the forces induced by the spin--gravity coupling
 vanishes, and thereby the top does follow a geodesic \cite{hojmanUFF}. Besides, in Ref.~\cite{hojmanUFF}, a different and more concrete experimental setting was proposed for tops moving ``parabolically'' in a non--geodesic orbit, where a measurement could be possibly performed.

This previous success in the agreement of the results which stem from Lagrangian theory for tops with experiment, leads to wonder what experimental settings can be appropriated to measure deviations of from geodesic orbits.  It is the purpose of this work to study the circular orbits of tops in a Schwarzschild background using this Lagrangian theory, showing how the spin induces non--geodesic motion of the test bodies. We apply these results to estimate these deviations for a possible circular orbit around Earth.
There exist other schemes that allow us to study the motion of tops in gravitational fields, such as post-Newtonian approximation for spinning massive particle \cite{will,barker2,larry1,larry2,porto1,porto2}. 
In particular, the chaotic motions in post-Newtonian systems of spinning compact binaries were investigated in  Refs.~\cite{jlevin,mdhartl,jlevin2,xwu,xwu2,syzhong,ywang,ghuang,lhuang,jluo}.
Also, circular orbits of tops on a Schwarzschild background have been studied using the MPE approach \cite{bini},
and the chaotic motions of tops in this spacetime background were explored in Refs.~\cite{ssuzuki,CVerhaare}.
However, in here we restrict ourselves to the Lagrangian formalism of Sec.~\ref{lagrsectifmr}, as it allows us to obtain an exact solution for the motion of tops in circular trajectories,  without the difficulties introduced by the MPE approach.

\section{Lagrangian theory for tops}
\label{lagrsectifmr}

The Lagrangian model for spinning particles consider tops with mass $m$, spin $J$, energy $E$
and total angular momentum $j$. The full theory is developed in Refs.~\cite{hojman1,hr,hojman2a,hojman2,hojman3,ha,hojmanUFF}, and we limit ourselves here to highlight its most relevant results.

\subsection{Equations of motion}

It is well-known that the velocity
$u^\mu$ of a spinning particle is not parallel, in general, to the canonical momentum vector $P^\mu$. The velocity vector may, under some circumstances, become spacelike \cite{hojman1,hr,hojman3}. However, the momentum vector remains always timelike and gives rise to the dynamical conservation law of mass $m^2\equiv P^\mu P_\mu> 0$ \cite{hojman1,ha}. The spin of tops is defined through an antisymmetric tensor $ S^{\mu \nu}$ (see below). The action $S=\int L\, d\lambda$ associated to the Lagrangian theory for tops is $\lambda$--reparametrization
invariant, where the Lagrangian $L (a_1, a_2, a_3, a_4) = (a_1)^{1/2}\mathcal{L} \left(a_2/a_1,
a_3/(a_1)^2, a_4/(a_1)^2\right)$
is an
arbitrary function of four invariants $a_1, a_2, a_3, a_4$, and
$\mathcal{L}$ is an arbitrary function of three variables where $a_1\equiv u^\mu u_\mu,\ a_2\equiv \sigma^{\mu \nu} \sigma_{\mu \nu}=
- \mbox{tr}({\sigma}^2),\ a_3\equiv u_\alpha \sigma^{\alpha \beta}
\sigma_{\beta \gamma} u^\gamma,\ a_4\equiv \mbox{det}({\sigma})$ \cite{hojman1,ha,hojmanUFF}, where
$u^\mu$ and $\sigma^{\mu \nu}$ are the top's velocity and angular velocity respectively defined in terms of derivatives with respect to the arbitrary parameter $\lambda$ (see Refs.~\cite{hojman1,hr,hojman2a,hojman2,hojman3,ha,hojmanUFF}).
The momentum vector $P_\mu$ and the antisymmetric spin tensor
$S_{\mu \nu}$ are canonically conjugated to the position and orientation of the top, $P_\mu \equiv {\partial L}/{\partial u^\mu}$ and $S_{\mu \nu} \equiv {\partial L}/{\partial \sigma^{\mu \nu}} = -
S_{\nu \mu}$.  Explicit examples of such Lagrangians can be found in Refs.~\cite{hr,ha}.
 In this way, it is found that the dynamics of a top describes a non--geodesic behavior, seen through the momentum equation  \cite{hojman1,ha,hojman2}
\begin{equation}
  \frac{D P^{\mu}}{D\lambda}\equiv\dot P^\mu+\Gamma^\mu_{\alpha\beta}P^\alpha u^\beta=-\frac{1}{2}{R^\mu}_{\nu\alpha\beta}u^\nu S^{\alpha\beta}\, ,
\label{momentummotion}
\end{equation}
and the equation for the spin tensor
\begin{equation}
  \frac{D S^{\mu\nu}}{D\lambda}\equiv\dot S^{\mu\nu}+\Gamma^\mu_{\alpha\beta}S^{\alpha\nu} u^\beta+\Gamma^\nu_{\alpha\beta}S^{\mu\alpha} u^\beta=P^\mu u^\nu-u^\mu P^\nu\, . \label{spinmotion}
\end{equation}
The overdot represents the derivative with respect to an arbitrary parameter ($\lambda$), in such a way that  velocity $u^\mu=\dot x^\mu$ is the derivative of coordinates. In addition, ${\Gamma^{\nu}}_{\rho \tau}$ are  the Christoffel symbols for the metric field $g_{\mu \nu}$ (the speed of light is set equal to $1$).
The six independent components of the antisymmetric spin tensor generate Lorentz transformations, and in order to restrict them to generate three dimensional rotations we impose the Tulczyjew constraint $S^{\mu \nu} P_\nu=0$ \cite{tulc,hojman1, hr,hojmanComment}. This constraint has been shown to be important in the consistency of a theory for spinning massive particles
\cite{hojmanComment}, as it can be deduced as a constraint which emerges from the Lagrangian of the theory, and not an external imposition on the top dynamics \cite{hr}  (there are other Hamiltonian formulations which do not require these constraints \cite{ambrosi}).
Lastly, in this theory, the (square) top spin $J^2\equiv \frac{1}{2} S^{\mu \nu}S_{\mu \nu}$ can be shown to be a conserved quantity \cite{ha,hojman1,hojman2a, hojman2,hojman3}.

The non--geodesic behavior of a top moving on a background gravitational field is determined by Eqs.~\eqref{momentummotion} and \eqref{spinmotion}, plus the constraint.  As a result,  the top can be interpreted as an extended object  that is subject to tidal forces due to gravity. Spin gives internal structure to the classical massive particles, and they cannot be longer described as pointlike objects. Due to the fact that  any extended object is crossed by infinitely many geodesics (only a pointlike object is  traversed by just one geodesic) the averaged motion does not align with any of the constituent geodesics, and the motion is, in general, non--geodesic. Similar effects have been studied for fields \cite{brdewitt,ahfields} (which are naturally extended objects) and electromagnetic waves \cite{brdewitt,ahEMwaves1,ahEMwaves2}.
Thus, one should expect that the inclusion of spin in the dynamics of massive particles should lead to non--geodesic orbits.

\subsection{Comparison with Mathisson-Papapetrou theory}

The above theory does not coincide with the MPE for massive spinning particles \cite{math,papa}. As it is discussed in Ref.~\cite{hojmanComment}, the MPE present several problems in their description for tops. The MPE formalism is composed by the third--order dynamical equations
\begin{equation}
 \frac{D P^{\mu}}{D s}=-\frac{1}{2}{R^\mu}_{\nu\alpha\beta}u^\nu S^{\alpha\beta}\, ,\quad
P^\mu= m u^\mu+u_\nu \frac{D S^{\mu\nu}}{D s}\, ,\quad
  \frac{D S^{\mu\nu}}{D s}+u^\mu u_\alpha \frac{D S^{\nu\alpha}}{D s}-u^\nu u_\alpha \frac{D S^{\mu\alpha}}{D s}=0\, .
\end{equation}
where $P^\mu$ is the momentum defined in that theory, and
$D/Ds$ is the $s$-parametrized covariant derivative for the proper time $s$ of the spinning particle. It is not difficult to obtain \cite{hojmanComment} that the system preserves the timelike behavior of the velocity $u_\mu u^\mu=1$. However, the MPE establishes that there is no dynamical conservation law for the mass \cite{hojmanComment}. In fact, it can be proved that the MPE, under the Pirani constraint $S^{\mu\nu}u_\nu=0$, the square of momentum $P_\mu P^\mu$ is not a constant of motion \cite{hojmanComment}.  On the other hand, under the Tulczyjew
constraint, it can be shown that the MPE implies that the mass is not a constant of motion \cite{hojmanComment}. Both results of the MPE represent a very undesired behavior for a relativistic theory of massive particles. We refer the reader to Ref.~\cite{hojmanComment}, where a deep discussion on the difficulties of the MPE formalism is presented.

Without a  Lagrangian formulation the canonical momentum and the spin tensor cannot be
appropriately defined. That is the reason the MPE equations have those difficulties, whereas the above Lagrangian theory does not.

\subsection{Integrability and non-integrability in Schwarzschild background }

Several authors have discussed the integrability and non-integrability of different models for spinning objects in a Schwarzschild field background.
For example, the non-integrability of system formed
by a  Schwarzschild black hole orbited by a spinning companion in the Mathisson-Papapetrou formalism under the extreme--mass--ratio limit was proved in Refs.~\cite{ssuzuki, CVerhaare}, showing the  chaoticity of the model.
Furthermore, in Ref.~\cite{jlevin2,jlevinprd} was found the onset of chaos in the post--Newtonian Lagrangian formulation of the two-black hole system with one body spinning, whereas  the post--Newtonian Hamiltonian formulation of the two-black hole system with one body spinning was shown to be integrable and regular \cite{konigs,konigs2}.
By using the construction of canonical conjugate spin variables  and the relation between Lagrangian and Hamiltonian approaches at the same post-Newtonian order \cite{wuxie,wuxie2}, it has been proved 
that these two models are integrable and non-chaotic, regardless of the post--Newtonian order \cite{wuxie3}.
On the other hand,  a system formed
by a  Schwarzschild black hole orbited by a spinning companion in the   extreme-mass-ratio limit, described by a Lagrangian theory, can be chaotic \cite{jlevin2}.

\section{Circular motion solution in a Schwarzschild background}

Several different general and exact solutions of the Lagrangian theory for tops have been found in Refs.~\cite{hojmanUFF, ha,hojman1,hojman2,hojman2a,hojman3,hk,zha,abk,ahkz}. Here we present only the key steps to obtain the solution for a circular motion of the top with spin perpendicular to the plane of motion.
 We refer to readers to those references for a full and detailed procedure to get the solutions for the equations of motion derived from the Lagrangian theory.

Assuming a Schwarzschild field background (describing approximatelly the Earth gravitational field), the equatorial motion of a top can be solved exactly, as any equatorial plane can be defined  for circular motion to take place. We write the metric in spherical coordinates $g_{tt}=1-2r_0/r$,
$g_{rr}=-\left(1-2r_0/r\right)^{-1}$, $g_{\theta\theta}=-r^2$,
$g_{\phi\phi}=-r^2\sin^2\theta$, where $r_0=GM$  with the gravitational constant $G$ and the Earth mass $M$. The circular motion is defined as such by $\dot r=0$. Besides, without any loss of generality, we can study the the motion in the
plane defined by $\theta=\pi/2$. If the top is initially in that
plane and $\dot\theta=0$, then it remains in that equatorial plane, where $P^\theta=0$
\cite{hojman1,ha,hojmanUFF}. In this solution, spin can be chosen to be orthogonal to the equatorial plane $S^{r\theta}=S^{\theta\phi}=S^{0\theta}=0$ \cite{hojman1,ha,hojmanUFF}, being parallel or antiparallel to the total angular momentum of the top along the whole trajectory. This total momentum angular $j$ for the top in this trayectory is a conserved quantity.
In  Refs.~\cite{hojman1,ha,hojmanUFF} is shown that
the general solutions for the momenta equations \eqref{momentummotion} are
$P_\phi=({-j\pm E J/m})/({1-\eta})$, and $P_t=[{E\mp j J r_0/(m r^3)}]/({1-\eta})$, with the dimensionless parameter $\eta={J^2 r_0}/({m^2 r^3})$. Here, the $\pm$ stands for two trajectories that depend on the spin orientation, parallel or antiparallel to the total angular momentum of the top, both of them remaining perpendicular to the plane of motion.
These two momenta are conserved
($\dot P_t =0$ and $\dot P_\phi =0$) for circular motion \cite{ha,hojmanUFF}, and thus, the E\"otv\"os  ratio is meaningless for this particular orbit. The non--geodesic motion manifests itself in changes of the velocity, not acceleration.

From the constant of motion $P_\mu P^\mu=m^2$, we get that $P^r=0= \left[{P_t^2}-\left({P_\phi^2}/{r^2}+m^2\right)\left(1-{2r_0}/{r}\right)\right]^{1/2}$, in consistency with the  circular motion solution, and the relation between the radial momentum and the radial velocity $\dot r=\left(1-{2 r_0}/{r}\right)\left({P^r}/{P_t}\right)=0$, given by solutions of the Lagrangian theory for tops \cite{hojman1,ha,hojmanUFF}. This constraint determines the energy of each trajectory
\begin{equation}\label{eqforenergyE}
{\left(E_\pm\mp \frac{j Jr_0}{mr^3}\right)^2}=\left(1-\frac{2r_0}{r}\right)\left[m^2{(1-\eta)^2}+\frac{1}{r^2}\left(-j\pm \frac{E_\pm J}{m}\right)^2 \right]\, ,
\end{equation}
of a  top  moving on a
circular orbit of radius $r$. 
The solutions of Eq.~\eqref{eqforenergyE}
for energy can be readily obtained as
\begin{equation}\label{eqforenergyE2}
E_\pm=\frac{\pm \Lambda_2+ \sqrt{\Lambda_2^2+\Lambda_1 \Lambda_3}}{\Lambda_1}\, ,
\end{equation}
where
\begin{eqnarray}
\Lambda_1&=&1-\frac{J^2}{m^2 r^2}\left(1-\frac{2 r_0}{r}\right)\, ,\nonumber\\
\Lambda_2&=&\frac{jJ}{m r^2}\left(\frac{3 r_0}{r}-1\right)\, ,\nonumber\\
\Lambda_3&=&\left(1-\frac{2 r_0}{r}\right)\left(\frac{j^2}{r^2}+m^2(1-\eta)^2\right)-\frac{j^2 J^2 r_0^2}{m^2 r^6}\, .
\end{eqnarray}
Circular orbits are obtained by the study of the behavior of the effective potential $V_\pm(r)=E_\pm(r)^2$ \cite{teukolsky}. By calculating the zeros of the derivative of the effective potential a condition for the radius of the orbits can be obtained,
 while their stability can be studied through the positive behavior of its second derivative. If spin is neglected ($\Lambda_1=1$ and $\Lambda_2=0$), the effective potential becomes simply $V=\left(1-{2 r_0}/{r}\right)\left({j^2}/{r^2}+m^2(1-\eta)^2\right)$, and from the zeros of the its derivative we obtain a relation between the allowed radius and
the angular momentum  of spinless particles \cite{teukolsky}
\begin{equation}\label{angmomenspinless}
j=\frac{m r \sqrt{r_0}}{\sqrt{r-3r_0}}\, .
\end{equation}
From where we obtain that circular orbits are allowed only for $r>3r_0$. 
By checking the second derivative of the effective potential, the stability of circular orbits are restricted only to $r> 6r_0$.
When spin is included, the condition for circular orbits can be obtained by the zeros of the first derivative of the complete solution for energies \eqref{eqforenergyE2}. The calculation is not straightforward, however for a small spin contribution $J\ll j$, and for tops far from the black hole $r_0\ll r$, the allowed stable circular orbits are still for $r\gg 6r_0$. This is the case for the estimations of the trayectory deviations for tops around Earth
studied in next section.

On the other hand, the  non--trivial spin evolution equations \eqref{spinmotion} relevant to the circular motion in the plane
$\theta=\pi/2$,
reduce to
${D S^{tr}}/{D\lambda}=0$ and ${D S^{t\phi}}/{D\lambda}=P^t \dot \phi-P^\phi$
 \cite{hojman1,ha,hojmanUFF}. These equations, together with the relations $S^{tr}=-{S^{\phi r}P_\phi}/{P_t}$ and $(S^{\phi r})^2={J^2 \left(P_t\right)^2}/({m^2 r^2})$ that can derived from the two constants of motion and the
Tulczyjew condition \cite{hojman1,ha,hojmanUFF}, allow us to get the angular velocity $\dot\phi_\pm$ for the two possible trajectories of this motion \cite{hojman1,ha,hojmanUFF}
\begin{equation}
 \dot\phi_\pm=\frac{1}{r^2}\left(1-\frac{2 r_0}{r}\right)\left(\frac{2\eta+1}{\eta-1}\right)\left(\frac{{-j\pm E_\pm J/m}}{{E_\pm\mp j J r_0/(m r^3)}}\right)\, ,
\label{phipunto}\end{equation}
where the energy $E_\pm$ are given by solutions of \eqref{eqforenergyE}.
Tops can have two tangential velocities $r\dot\phi_\pm$, according to Eq.~\eqref{phipunto}.
The interplay of its spin with gravity, introduces different corrections in this tangential velocity, which depends on the spin orientation, such that the top with antiparallel spin is faster than the one with the parallel spin to the total angular momentum $r\dot\phi_->r\dot\phi_+$.

A possible measurement of a  maximal manifestation of non--geodesic motion can be achieved if two test bodies (with equal composition) are set to rotate in order to have opposite (internal) angular momenta directions, parallel and antiparallel to the total angular momentum of the circular motion of the satellite. In such cases, any deviation from geodesic orbits must be reflected in different measurements of the angular velocities of the test bodies. Thus, the dimensionless ratio
\begin{eqnarray}\label{deltaparameter}
\delta=\frac{\dot\phi_--\dot\phi_+}{\dot\phi_-+\dot\phi_+}=\frac{\left[-j+\frac{E_+J}{m}\right]\left[E_-+\frac{jJr_0}{mr^3}\right]+\left[j+\frac{E_-J}{m}\right]\left[E_+-\frac{jJr_0}{mr^3}\right]}{{\left[j-\frac{E_+J}{m}\right]\left[E_-+\frac{jJr_0}{mr^3}\right]+\left[j+\frac{E_-J}{m}\right]\left[E_+-\frac{jJr_0}{mr^3}\right]}}\, ,
\end{eqnarray}
is non--zero in the case of circular orbit for the non--geodesic  behavior of tops.

Notice that for these circular motions described by solution \eqref{phipunto}, tops do not present tangential accelerations $\ddot\phi=0$, as the radius remains constant. Therefore, there is no relative acceleration between the two test bodies.
Furthermore, when spin is neglected $J=0$ ($\eta=0$), a massive particle can only have a unique angular velocity $\dot\phi=j/(r^2 E)$ and an unique energy,
 yielding the usual result $\delta=0$ for geodesic motion in the Schwarzschild field \cite{hartle}. The approximately vanishing  E\"otv\"os  ratio and $\delta=0$ are the results measured in the MS mission \cite{microscope}, which is in agreement with the Lagrangian theory for tops.

\section{Estimations of the trajectory deviations}

The inclusion of spin into the the test bodies is essential for experiments carried out to demonstrate the validity of non--geodesic motion of massive spinning bodies.   To study the first order corrections to the circular orbit of test bodies due to its spin, let us consider the following approximations. Let us study  tops motion with $J\ll j$.  In this case, the total momentum angular $j$ is approximately by the orbital momentum angular of the top.

At first order in spin, the energy solutions of Eq.~\eqref{eqforenergyE} is
\begin{equation}\label{energyatfirstorder}
{E_\pm}\approx m\sqrt{\left(1-\frac{2r_0}{r}\right)\left(1+\frac{j^2}{m^2r^2} \right)}\mp \frac{jJ}{mr^2}\left(1-\frac{3 r_0}{r}\right)\, .
\end{equation}
With this solution, we can obtain the angular velocity \eqref{phipunto} of tops at first order in spin to be
\begin{equation}\label{phiyatfirstorder}
\dot\phi_\pm\approx \frac{1}{mr^2}\sqrt{\frac{1-\frac{2r_0}{r}}{1+\frac{j^2}{m^2r^2}}}\left[j\mp J \sqrt{\frac{1-\frac{2r_0}{r}}{1+\frac{j^2}{m^2r^2}}}\right]\, .
\end{equation}
Therefore, at first order in spin, the $\delta$-parameter \eqref{deltaparameter} becomes
\begin{equation}\label{deltatfirstorder}
\delta\approx \frac{J}{j}\sqrt{\frac{1-\frac{2r_0}{r}}{1+\frac{j^2}{m^2r^2}}}
\end{equation}

Now, consider a possible experiment near the Earth surface (where $r_0\sim 4.4\times 10^{-3} \mbox{[m]}\ll r\sim 7\times 10^6\mbox{[m]}$) in order to measure the non-geodesic behavior of a top in circular trayectory.
This kind of experiment  have been perfomed by the MS mission \cite{microscope}.
Let us consider a model of two test bodies with the same total angular momentum (which is a conserved quantity for each one) but different spin orientations. Thus, considering a small spin contribution and $r_0\ll r$,  the top total angular momentum  can be approximated by Eq.~\eqref{angmomenspinless}. Notice that this angular momentum is consistent with Kepler's law of motion, as for a general angular momentum
 $j= m r v$ (with test top's velocity $v$ much smaller than the speed of light) we can write $v=r\Omega$ for a circular orbit, where $\Omega$ is the angular frequency of the satellite's trajectory. By Kepler's law, this frequency is related to the trajectory radius by $\Omega^2={r_0}/{r^3}$. Using this, the angular momentum becomes $j\approx m \sqrt{r r_0}$, which is an approximation to Eq.~ \eqref{angmomenspinless}.

\begin{figure}
\includegraphics[width=0.5\textwidth]{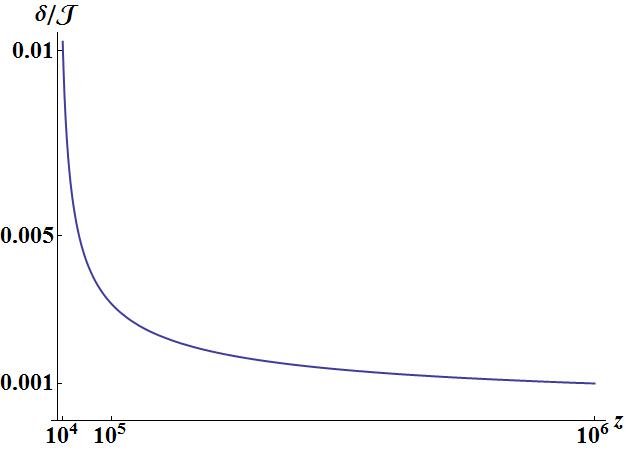}
\caption{Behavior of the parameter \eqref{deltatfirstorder2}, normalized to ${\cal J}=J/(m r_0)$, in terms of dimensionless distance $z=r/r_0$. The plots are for $z\gg 1$, according to our approximations.}
\label{fig1}
\end{figure}

Thus, the energy of top \eqref{energyatfirstorder}  at first order becomes
\begin{equation}\label{energyatfirstorder2}
{E_\pm}\approx m\left(1-\frac{r_0}{2r}\right)\mp \frac{J}{r}\sqrt{\frac{r_0}{r}}\left(1-\frac{3 r_0}{2r}\right)\, ,
\end{equation}
while  the approximated angular velocity \eqref{phiyatfirstorder} at first order is
\begin{equation}
\dot\phi_\pm\approx \frac{1}{r}\sqrt{\frac{r_0}{r}}\left(1-\frac{3 r_0}{2r}\right)\left[1+\frac{3r_0}{2r}\mp \frac{J}{m\sqrt{r_0 r}}\left(1-\frac{3 r_0}{2r}\right)\right]\, .
\end{equation}
Notice that the gravitational correction factor $3r_0/2r$ to the spin-coupling  also appears in post-Newtonian theories \cite{schafer}.
 Lastly,  the parameter \eqref{deltatfirstorder} at first order becomes simply
\begin{equation}\label{deltatfirstorder2}
\delta\approx \frac{J}{m\sqrt{r_0 r}}\left(1-\frac{3r_0}{r}\right)\, .
\end{equation}
 The behavior of this parameter is shown in Fig.~\ref{fig1}, in terms of dimensionless quantities. We have  plotted $\delta/{\cal J}$ as  function of the normalized radius $z=r/r_0$, where ${\cal J}=J/(m r_0)$. The plot is presented for $z\gg 1$, according to the used approximations.
 It is straightforward to show from \eqref{deltatfirstorder2} that $\delta/{\cal J}=(1-3/z)/\sqrt{z}$, implying that the parameter has the same behavior for any given spin,  depending only on the orbit radius $r$.
From the plot, we can see that $\delta$ decays as $1/\sqrt{z}$.

We can estimate in a closer form the parameter \eqref{deltatfirstorder2} for a test spinning body.
Assuming that the intrinsic spin of each top can be  estimated as $J\sim m d^2\omega$, where $d$ is a characteristic length of the experimental test top body and $\omega$ is its internal angular frequency of rotation.
 Thus, the correction to the circular trajectory of two massive test objects with different spin orientations can be estimated through the ratio \eqref{deltatfirstorder2} to be $\delta\approx ({d^2\omega}/{\sqrt{r_0 r}})(1-3r_0/r)$.
This approximated $\delta$--ratio depends on the the test top parameters $r$, $\omega$ and $d$.
Assuming that the test tops have dimensions of the order of $d\sim 10^{-2}$[m], then the ratio gives of the order
\begin{equation}\label{eotvostraj2}
\delta\approx 10^{-15} \ \omega \, .
\end{equation}
where the internal angular frequency of tops is measured in [Hz].  For a higher intrinsic angular velocity, a larger deviation from a geodesic path can be achieved.

\section{Conclusions}

The main purpose of this work is to bring attention to the notion of non--geodesic motion, showing that it is the spin (internal dynamics) and not the  composition, size or shape of the test bodies, the key for detecting a possible non--geodesic trajectory.

In the simplest case of a circular motion for a spinning massive particle, the complete dynamics can be solved exactly, and it describes a non-geodesics motion whose deviations depend on the magnitude and direction of the particle's spin. Notice that the solution presented above has constant momenta, and thus, the non--geodesic motion cannot be detected by measuring the E\"otv\"os  ratio, as the non--geodesic motion is manifested only in the change of the velocities of test bodies. Of course, this is not a general rule, and more complicated motions in different spacetimes may indeed present non--vanishing E\"otv\"os  ratios \cite{ha,hojmanUFF,zha}. As the $\delta$-ratio show in Eqs.~\eqref{deltaparameter}, when spin is negligible then geodesic motion is expected. Therefore, it comes as no surprise that the EP is repeatedly confirmed for non-spinning test bodies.

Any experimental setup designed to measured non--geodesic motion must be constructed in order to capture the spin-gravity coupling and its effects. In particular, for experiments around Earth the $\delta$--ratio has enough freedom to adjust the parameters of the orbit radius of the satellite and the characteristic length and inner angular velocity of test body,
depending on the accuracy of the setup.
 Thus, any  experimental setting should also consider the angular momenta of test bodies in order to prove the validity of non--geodesic motion due to spin.

\end{document}